\documentclass[sigconf]{acmart}
\usepackage{epsfig,endnotes}
\usepackage{pifont,hyperref,color,graphicx,dirtree,multirow}
\usepackage{dblfloatfix} %
\usepackage{adjustbox}

\newcommand\chk{\color{black}{\ding{51}}}
\newcommand\crs{\color{red}{\ding{55}}}

\newcommand\sqr[1]{{\color{#1}{\ding{110}}}}

\newcommand\naa{-}

\usepackage{enumitem}
\usepackage{setspace}
\usepackage{subcaption}
\setlist[itemize]{leftmargin=3mm}

\newcommand{\one}{{\em (i)}}
\newcommand{\two}{{\em (ii)}}
\newcommand{\three}{{\em (iii)}}

\definecolor{undisturbed}{HTML}{005A00}
\definecolor{degraded}{HTML}{649B23}
\definecolor{deforested}{HTML}{FF870F}
\definecolor{regrowth}{HTML}{D2FA3C}
\definecolor{water}{HTML}{008CBE}
\definecolor{other}{HTML}{C8C8C8}

\renewcommand\footnotetextcopyrightpermission[1]{} %

\begin{document}

\date{}

\title{Planetary computing for data-driven\\ environmental policy-making}

\author{Patrick Ferris, Michael Dales, Sadiq Jaffer, Amelia Holcomb, Eleanor Toye Scott, \\Thomas Swinfield, Alison Eyres, Andrew Balmford, David Coomes,\\ Srinivasan Keshav, Anil Madhavapeddy}
\affiliation{%
  \institution{Departments of Computer Science \& Technology, Plant Sciences and Zoology\\Cambridge Conservation Initiative}
  \city{Cambridge}
  \country{UK}
}

\maketitle
\renewcommand{\shortauthors}{Ferris, Dales, Jaffer, Madhavapeddy, et al}
\pagestyle{plain} %

\thispagestyle{empty}

\subsection*{Abstract}
We make a case for {\em planetary computing} -- infrastructure to handle the ingestion, transformation, analysis and publication of global data products for furthering environmental science and enabling better informed policy-making. We draw on our experiences as a team of computer scientists working with environmental scientists on forest carbon and biodiversity preservation, and classify existing solutions by their flexibility in scalably processing geospatial data, and also how well they support building trust in the results via traceability and reproducibility. We identify research gaps in the intersection of computing and environmental science around how to handle continuously changing datasets that are often collected across decades and require careful access control rather than being fully open access.

\section{Introduction}

Policies designed to tackle rising CO$_2$ emissions~\cite{tollefson_carbon_2022}, rapid biodiversity loss~\cite{tollefson_humans_2019}, and desertification~\cite{hu_northward_2022} are becoming increasingly data-driven. For example, satellite and drone remote sensing~\cite{dubayah2020global, tang_drone_2015}, wireless, ground-based sensor measurements~\cite{burgess_harnessing_2010} and climate modelling~\cite{jakob_need_2023} all play important roles in modern climate research and policy formulation. Processing and analysing a diverse range of datasets using computational pipelines has allowed rapid progress within ecology~\cite{mutanga2019google}.
However, the \textit{computer systems} required to effectively ingest, clean, collate, process, explore, archive, and derive policy decisions from the raw data are presently not usable by non-CS-experts, not reliable enough for scientific and political decision-making, and not widely and openly available to all interested parties across the globe.

We thus make the case for {\em planetary computing}: the digital infrastructure required to handle the ingestion, transformation, analysis and publication of global data products for furthering environmental science. Planetary computing systems have a vital role to play in not only powering the distillation of insights about our world, but also in building public trust in the resulting policy actions by enforcing standards of transparency, reproducibility, accountability and timeliness in the decision-making process.

While planetary computing is superficially similar to classic big data processing, it presents some unique challenges that we have learnt first-hand. We will first describe the global ecological analyses that our joint team of computer and environmental scientists have built in the past three years (\S\ref{s:motive}), and the lessons learnt (\S\ref{s:lessons}) from that implementation process.  We distill some common requirements for planetary computing (\S\ref{s:requirements}) and find that existing solutions only partially solve the computer systems problems (\S\ref{s:gap}).
We then discuss the unsolved computing research opportunities offered by planetary computing and discuss the pros and cons of building systems intended to be used by environmental scientists and policymakers (\S\ref{s:discuss}).
Our ultimate aim is to grow an ecosystem that spans organisations (ranging from NGOs to governments to corporations who all need access to intelligence about planetary health), that can survive the failure of any one controlling entity in the coming decades, and that supports transparent and reproducible computational science whilst respecting the sensitivity of the underlying observational datasets.

\subsection{Motivating Environmental Scenarios}
\label{s:motive}

At the Cambridge Conservation Initiative, we have had a joint team of computer scientists and ecologists working together for several years to tackle the following environmental challenges, which we will use to motivate the need for planetary computing.

\paragraph{Analysing effectiveness of rainforest protections}
There are many worldwide conservation projects across the equatorial belt protecting millions of hectares of old-growth tropical rainforest, resulting in reduced CO$_2$ emissions from deforestation (an approach known as REDD+~\cite{angelsen2009realising}). Our challenge is to calculate the ``additionality'' of these interventions such that we can quantify how many tonnes of CO$_2$ emissions have been avoided per year. The interventions typically take the form of sustainable livelihoods (e.g., mixed-forest cocoa plantations) that are an alternative to clearcutting the forest~\cite{mbomaalternative}, and the CO$_2$ additionality measurement is vital towards justifying financing the intervention. The current measurement mechanism is for projects to self-certify the background deforestation rates in the region, and then manually measure the net gain in above-ground biomass \textit{vs} their baseline. However, this approach often overestimates the avoided emissions and results are not comparable across projects with differing baselines~\cite{doi:10.1126/science.adh3426}. Instead, our computational solution uses global satellite data from the past twenty years to process worldwide estimates of forest biomass~\cite{dubayah2020global,achard2009monitoring} and apply statistical counterfactual analysis~\cite{tmfv2} to comparably identify the additionality resulting from the interventions. The results are adjusted for impermanence~\cite{swinfield2023realising} and converted into high-quality carbon credits that are used to offset unavoidable emissions such as international air travel with comparable climate benefit from the avoided forestry emissions~\cite{cowg}.

\begin{figure*}[h]
\centering
\begin{subfigure}{.5\textwidth}
  \centering
  \includegraphics[width=.95\linewidth]{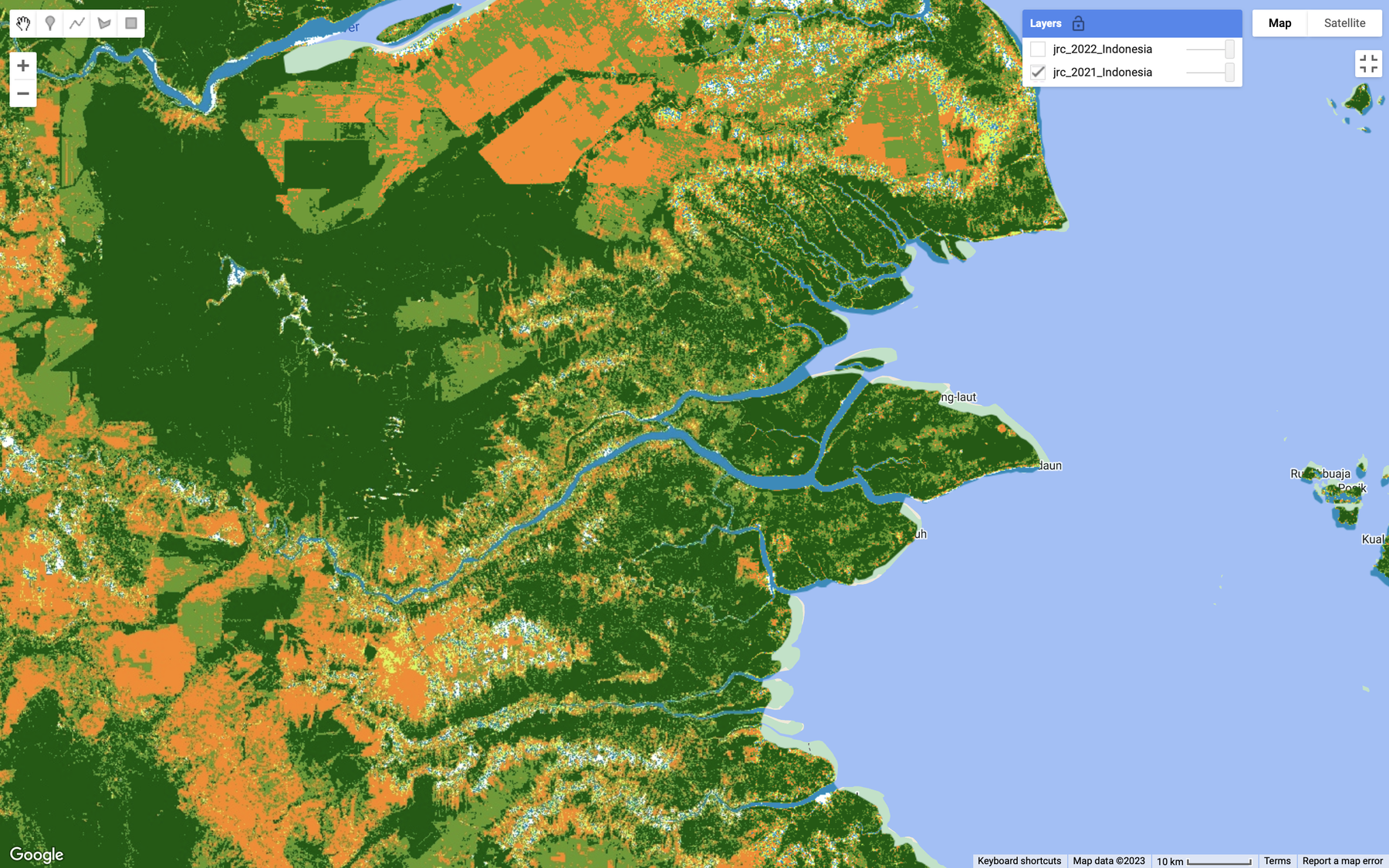}
\end{subfigure}%
\begin{subfigure}{.5\textwidth}
  \centering
  \includegraphics[width=.95\linewidth]{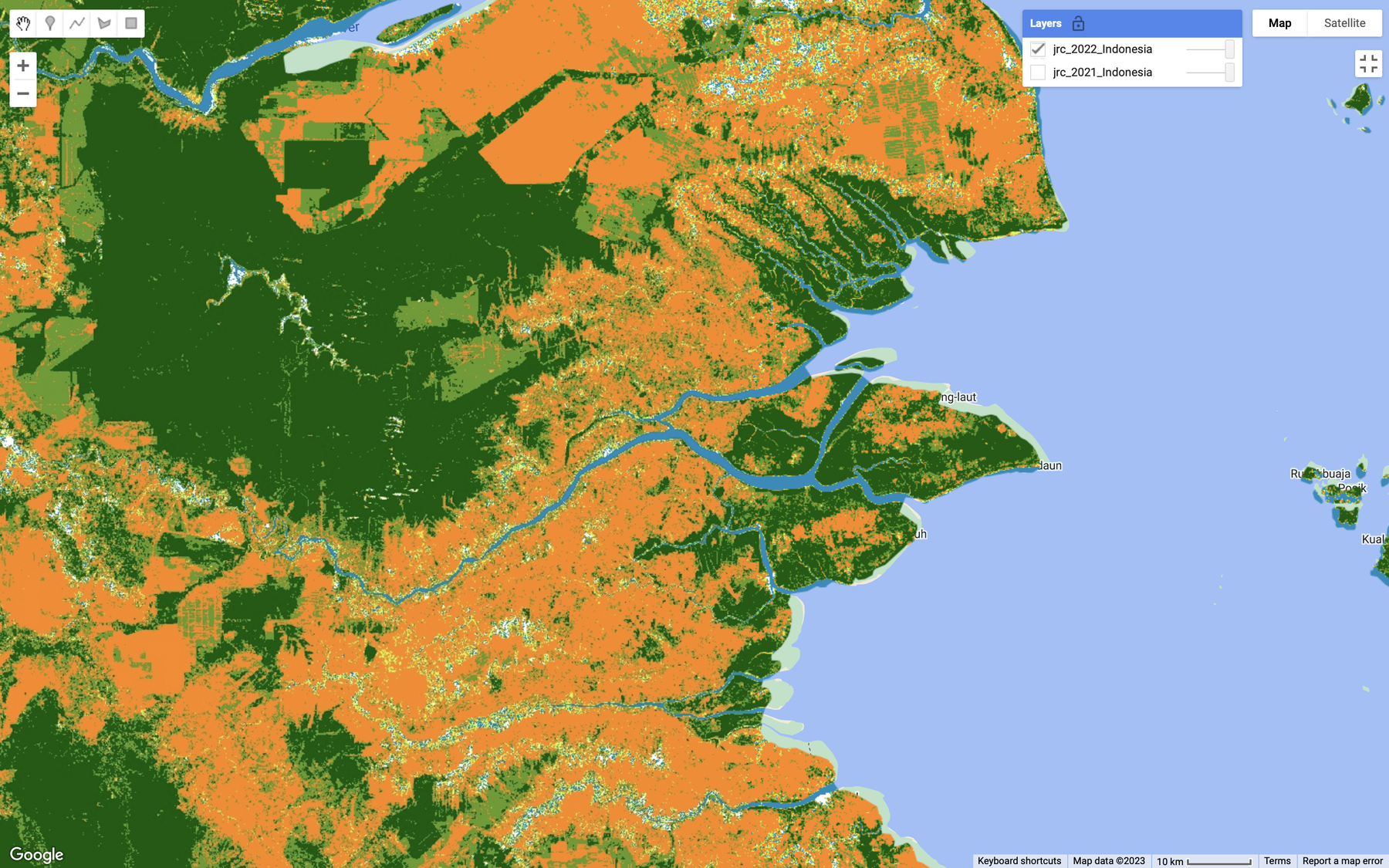}
\end{subfigure}
\caption{Two versions of the JRC TMF dataset showing the same land use class data for an area of Indonesia in \emph{2008}. The left image shows the \emph{2021} data and the image on the right shows \emph{2022} data. \sqr{undisturbed} undisturbed, \sqr{degraded} degraded, \sqr{deforested} deforested, \sqr{regrowth} regrowth, \sqr{water} water, \sqr{other} other}
\label{fig:jrc-diff-vis}
 \vspace{-1em}
\end{figure*}

\paragraph{Calculating worldwide extinction rates}

Ecologists assess areas of habitat data to generate worldwide extinction statistics~\cite{brooks2019measuring}, but must not reveal individual observation points or else species may come under threat from poachers~\cite{lindenmayer_not_2017}. To generate this aggregate data they combine satellite data (Landsat~\cite{wulder2022fifty}, MODIS~\cite{justice2002overview}, Copernicus~\cite{buchwitz2018copernicus}, GEDI~\cite{dubayah2022gedi}) with readings collected manually over decades~\cite{harfoot_using_2021}. The data is highly variable in quality and requires cleaning and normalisation~\cite{HOLCOMB2023100106}, before machine learning is used to train models to interpolate missing data. Information derived from the data is used to direct habitat regeneration and protection efforts, but must be regenerated monthly as new data arrives.  When challenged it should be possible to reveal the provenance of conclusions to auditors, even from decades-old observations.  The aggregate (around 92 petabytes of data) is processed across 29,000+ terrestrial species and used to create a global metric for extinction rates known as ``LIFE''~\cite{eyres_life_2023}. The metric is published as a versioned set of raster maps ($\approx$500 GB of final results) for general use, but ecologists also need access to individual species data.

\paragraph{Land-use policy for biodiversity preservation}
Food and fibre production trades off against natural habitats, and understanding where to do this requires jurisdictional land management~\cite{balmford_how_2019}.
A civil servant assessing different methods of evaluating the impact of land use changes on biodiversity needs to access datasets for their country that have a reasonable resolution ($<$100 metres/pixel and so 150GB/layer storage needed), across all the species on the IUCN extinction list (10000+ entries~\cite{harfoot_using_2021}), and go back 30 years. Similarly, natural resource managers rely on being able to work on zoomed-out/cropped data for interactive and iterative exploration of potential land use policies, and then scale to cluster compute levels for a country-wide run.

\section{Lessons from the Field}
\label{s:lessons}

When we began these projects, building the computer systems to solve the environmental challenges looked straightforward. The amount of aggregate satellite data was around a dozen terabytes, and so even a single many-core machine was able to run the data processing pipeline involved~\cite{mcsherry2015scalability}. However, as the computer scientists delved deeper, the subtle difficulties of handling ecological analyses started to come to light. Here we highlight four ways in which applying computation to ecology has scaled up uncertainty in the results produced.

\subsection{Uncertainty in Data}
\label{section:data}

The first task we had was to get a stable set of earth observation data on which to build our analyses, and we quickly discovered just how many variations there are of seemingly similar observations.
For example, we used the Tropical Moist Forest (TMF) dataset by the European Commission's
Joint Research Centre (JRC)~\cite{tmf} that calculates land use class (LUC) data worldwide from satellite observations and has been cited over 200 times.\footnote{According the Google Scholar as of 2023-12-14} TMF historic data is upgraded as new algorithms and analysis become available (which is good practice), but 
the means by which the data is published does not make these updates to historically available data obvious. Once updated, earlier versions on which other calculations (and papers published) were made are no longer easily available, hugely impacting research reproducibility.

The extent to which these differences matter is highly dependent on where in the world the analysis is being carried out. For our work, we looked at the tropical forest belt in Indonesia and found large differences when aggregating the LUC data across the country. When comparing two versions of the data (one from 2021 and the other from 2022) for the {\em same} historical year (2008), we see an increase of 3.05\% in land classed as \emph{deforested} and a decrease in land classed as \emph{undisturbed} of 0.59\% and \emph{degraded} of 2.00\%\footnote{Analysis and source code: \url{https://github.com/quantifyearth/jrc-diff}}. Whilst seemingly small, these numbers should be taken in the context of the area of Indonesia (approximately 1.9 million squared kilometres). The change in amount of land incorrectly classed as deforested is approximately the size of Togo! Figure~\ref{fig:jrc-diff-vis} is a visualisation of these changes; differences are partially caused by ``improvements and corrections of errors in the Annual Change collection in the sequence of values for deforestation of old regrowth forest...''~\cite{jrc-update}.
Without access to the original 2021 JRC release, these differences propagate silently through further downstream research. We observed similar changeable datasets in other satellite outputs~\cite{dubayah2022gedi}.

Perhaps counter-intuitively, uncertainty in data is at times a desirable property: certain datasets are highly sensitive and can result in the slaughter of species if accidentally released. Wildlife tracking by sensors, for example, can reveal animal locations and that can help people capture, harm or kill tagged animals~\cite{animal-tracking,tigerpoach}. There have been cases of tracking datasets being petitioned to be opened to assist with (often illegal) hunting~\cite{grover2001one}. To help resolve this tension, ecologists have published extensive decision trees to guide the choice of whether to open a dataset or not~\cite{tulloch2018decision}.  The outcomes range from restricting data by progressively masking species identity, the high resolution maps of their locations, and aggregating results. There is also the ethical conundrum of data gathered causing adverse impact on the local communities. Remote sensing technology can disturb existing collective responsibility~\cite{slough_satellite-based_2021} and potentially result in adverse livelihood impacts that increase pressure to resort to unsustainable approaches (such as clearcutting forests) to make up for the lost income~\cite{doi:10.1177/19400829211014740}.  These factors  resulted in reluctance from our conservation partners to publish their datasets on conventional hosting platforms, and instead manually manage it.

\subsection{Uncertainty in Code}

Ecologists have therefore become de facto data-scientists, requiring that they not only understand their own subject domain, but also they have to learn to work with unwieldy datasets, HPC computing infrastructure, and to write code in languages that are at best aimed at data-science (R or Julia) or general purpose languages that have had support for data-science grafted on to them (Python with libraries such as Numpy, Pandas, and so forth), with the big caveat that none of these languages are designed for petabyte-scale computation. It took software engineering, where the main output of the discipline is code, many years of producing code and many public failures in software quality to get to grips with now common practices such as version control, testing, linting, so it should come as no surprise that those tools have yet to make their way into other domains where code is seen as a means rather than as an end.

We, as others have done~\cite{ecocode, lowavail}, observed this first hand as we collaborated with our ecologist colleagues, that scripts get passed around, tweaked repeatedly, and then cast aside once the result has been produced. This is partly because the tools produced -- such as the ubiquitous \texttt{git} -- are challenging enough for experts to fully utilise correctly. But the architecture of these tools also does not acknowledge the nature of the work of these scientists: the tools encourage respect for the final tried-and-tested code, whereas data science is often an exploratory and iterative process where most versions of the code are not seen as correct and thus not worthy of committing, and then when that final version does emerge, the author is straight onto the next challenge with their newly minted data. Some form of automatic versioning of code {\em and} data would benefit exploratory programming~\cite{mashup-vcs}.

The motivation to do better is now emerging as scientific journals now often require the public availability of code~\cite{barnes_publish_2010} alongside the published manuscript, but that is often performed as a last step when the coupling between the published data and the specific version of the code used to generate it as been weakened. Software engineering practices such as Continuous Integration (CI) are not being used as a way to tie the output artefacts to the code that created it, partly due to the need to master yet another specification language, but also due to the difficulties of exporting large datasets from a local  development environment. But the need for a tight coupling between results and code was clear to the team.

Whilst in CS there is an awareness of formal methods to assess whether code does what it is meant to, or at least a strong suite of unit tests to give that confidence, that was not the case for those working with these large datasets, again due to the exploratory nature of the work until publication. The lack of systematic testing were of concern to the whole team since simple programming errors could easily lead to the need to retract publications~\cite{merali_computational_2010}. The dataset scale meant that visualisation tools did not make it easy to check intermediate results when processing tens of thousands of species in maps at <100m resolution.

\begin{figure}
    \centering
    \includegraphics[width=0.7\linewidth]{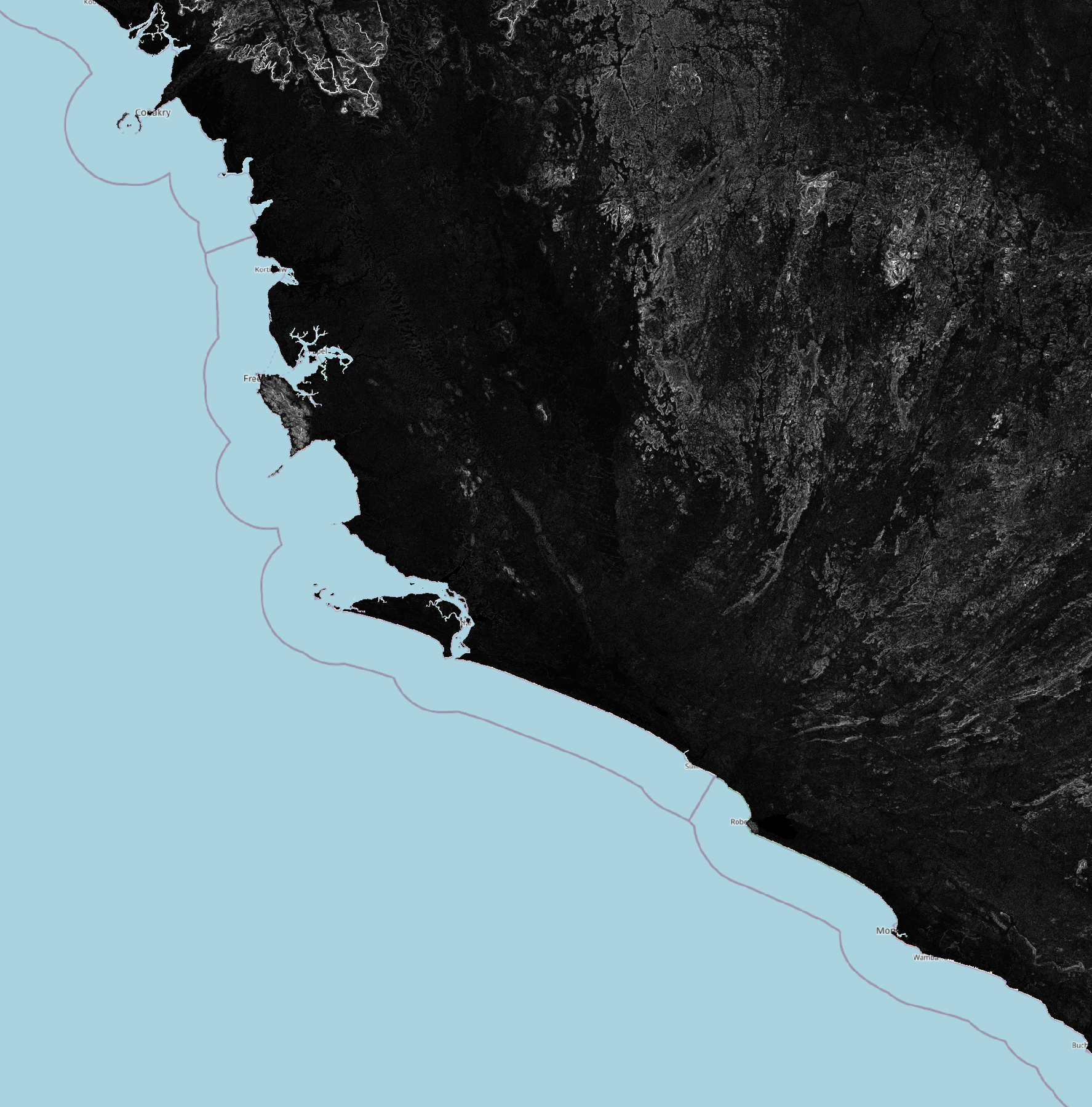}
    \caption{Showing the difference in terrain ruggedness index (TRI) as calculated by the \texttt{gdaldem tri} command between GDAL versions 3.2 and 3.3. The difference would ideally be zero (all black).}
    \label{fig:gdal-tri}
    \vspace{-1em}
\end{figure}

\subsection{Uncertainty in Dependencies}

Even when care is taken, without reproducible computing environments (e.g., using a system like Nix~\cite{nix}), the sheer multiplicity of factors that could change the final results was unmanageable within conventional operating systems. For example, consider the popular geospatial data manipulation library GDAL~\cite{gdal}. Figure~\ref{fig:gdal-tri} shows the difference between data derived with \emph{the same} command-line for calculating a terrain ruggedness index (TRI) but with GDAL 3.2 and GDAL 3.3.\footnote{Source code: \url{https://github.com/quantifyearth/gdal-tri-diff}} The difference is due to changes in the default algorithm that were necessary, but has introduced pixels differing by as much 372m in this example.

Similarly, subtle changes in hardware versions led to changes in results when the same pipeline was run on different machines. While processing biodiversity maps, for example, we observed unexpected small increases in habitat removal scenarios (where there should only be species loss). Tracing this led down a rabbithole which  ended with a rounding discrepancy between array and scalar operations on specific new versions of the AMD EPYC CPUs; they added 512-bit SIMD instructions that had different rounding properties to narrower SIMD and scalar operations. In isolation the rounding errors are insignificant (below the CPU floating point epsilon value), but when summed across thousands of species started to have significant impact. All the biodiversity scenarios were affected by the error, but it was hidden in most scenarios where loss and gain is expected. The sheer volumes of data made it hard to inspect intermediary results in fine detail, and it was fortunate that we did a full loss scenario that flagged the discrepancy -- and the team had enough computer science expertise to trace the right root cause across software and hardware and discuss it upstream.\footnote{See \url{https://github.com/numpy/numpy/issues/25269}}

The uncertainty in dependencies extended to popular platforms for doing geospatial analysis like Google Earth Engine (GEE) -- an all-in-one, code and data platform for doing ``planetary-scale analysis'' ~\cite{gee}. It has had widespread positive impact and enabled many ecologists to produce useful analysis that would have been difficult to achieve otherwise. One example is the global map of travel time to healthcare facilities which can be used as a proxy for ``remoteness'' in ecological analysis \cite{weiss_global_2020}. 
However, we started running into the limitations of GEE being a closed-source platform; for example, we could not adequately specify specific historic datasets (such as the aforementioned JRC TMF dataset) in a way that was guaranteed to be immutable and therefore reproducible. Instead, we relied on opaque identifiers that are potentially mutable, and on an open platform this behaviour could have been verified.

This lead onto concerns about the longevity of the computation platform to which the conservation research work was tied. It is typical for some interventions to be planned for decades (trees live a long time!), and involve thousands of people and dozens of organisations spread geographically. A platform like GEE -- no matter how effective in the short-term -- is tied to a single large corporation that has a record of closing down non-core products~\cite{stadia}. Many datasets in GEE have also been processed to make them easier to consume, but the methodology by which they have been processed is not published, preventing easy migration away from GEE once datasets are adapted.
Our use of GEE was thus a mixed bag -- it is extremely effective to use for interactive scientific exploration of large-scale datasets, but is not a platform that we could extend to satisfy our requirements for scientific reproducibility (key to any comprehensive peer-review process) and long-term commitments to ensuring our results can be explained in decades to come.

\subsection{Uncertainty in Policy}
\label{section:uncertain-policy}

\begin{figure}
    \centering
    \includegraphics[width=0.99\linewidth]{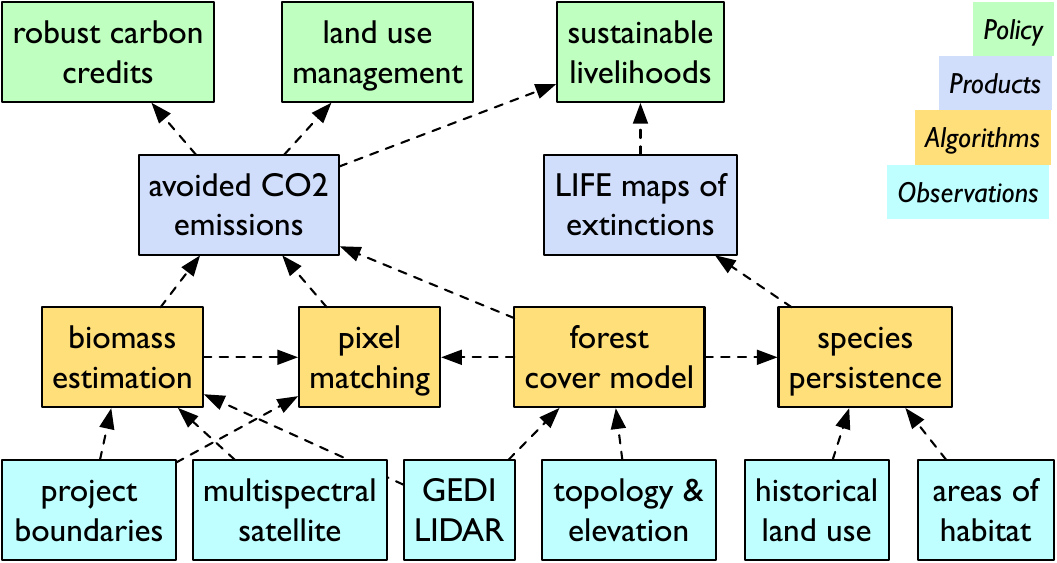}
    \caption{An illustrative pipeline for two of the motives in \S\ref{s:motive} and the policy scenarios that could be derived from the results. For a fuller specification see~\cite{tmfv2} or ~\cite{eyres_life_2023}.}
    \label{fig:pos-before}
    \vspace{-1em}
\end{figure}

Policy derived from poor data is not new; e.g. Herndon et al. found selective exclusion of data, spreadsheet coding errors and inappropriate statistical methods in Reinhart and Rogoff's paper on the links between high public debt and economic growth, a paper which was highly cited by influential statespeople~\cite{herndon-spreadsheet}. But whilst environmental science has dealt with uncertainty for a long time; e.g.,~sampling errors in biodiversity analysis~\cite{freckleton_census_2006}, a scientific method which incorporates large datasets and multiple stages of computation can compound the methodological uncertainty that already exists in the field.

Figure \ref{fig:pos-before} illustrates how this happens for the ecological challenges (\S\ref{s:motive}) we are tackling. The computational pipeline needs to additionally account for uncertainty across the source data, of varying algorithms, and of tracking the data products that are used by policymakers when they enact environmental change.
We thus next explore how planetary computing can build trust in decision-making derived from the results of global data analysis, and the considerations of a system that is sympathetic to the needs of policymakers and those that wish to challenge the proposed policies.

\section{Scoping planetary computing}
\label{s:requirements}

\begin{figure*}[t]
\includegraphics[width=1.8\columnwidth]{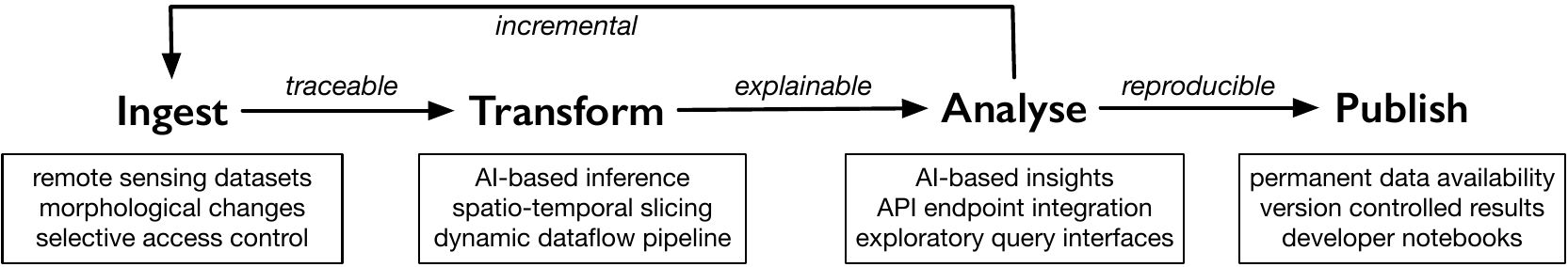}
\caption{Ideal dataflow pipeline for a planetary computing engine}
\label{f:ark}
\end{figure*}

In order to urgently accelerate our actions towards addressing the climate crisis, we advocate for a system that focuses on supporting environmental scientists and policymakers as its primary users, enabling collaboration on data-driven scenarios (\S\ref{s:motive}). %

\subsection{Capabilities}

Environmental scientists need to access \textbf{large-scale input datasets} consisting of:
\one{} primary observation data from satellites~\cite{dubayah2020global} that is petabyte-scale or direct ground measurements~\cite{nugent2018inaturalist, trolliet2014use};
\two {} derived sources from algorithmic transformation or AI-based inference; and
\three{} previous results derived by third parties or from earlier runs.
Programmers then \textbf{describe computation} over these datasets that:
\one{} is either algorithmic or machine learning-based, using a mix of CPUs and GPUs;
\two{} needs to autoscale to permit local development followed by global analysis;
and
\three{} can be expressed by a non-CS expert, ideally with a visual interface~\cite{9833110}, in a language like R or Python, or even with large language model assistance~\cite{sarkar2022like}.
The \textbf{derived products} provide higher level insights and must be archived for the long term, while:
\one{} tracking provenance on input data and enforcing privacy constraints on output results;
\two{} being independently verifiable when given access to the source data;
and
\three{} incrementally recalculating to allow for interactive exploration and incorporation of local data.
These requirements translate to the data pipeline shown in Figure~\ref{f:ark}, with four phases: 
\begin{itemize}
\item {\bf Ingestion}: is the acquisition of remote sensing datasets. Publishers often serve them via adhoc HTTP/FTP servers, requiring polling download scripts. The data formats vary (e.g., GeoJSON, TIFFs) and typically need to be normalised into a format such as a spatially indexed columnar store due to their size. There is rarely systematic version control from the data providers (\S\ref{section:data}). 
\item {\bf Transform}: is the dynamic computation pipeline over the very large datasets {\em at scale}, usually expressed in multiple languages (commonly Python, R, Julia or Fortran). The algorithms must scale to allow the ecologist to experiment locally before transferring their work to an HPC cluster. Additionally, machine learning is often used to interpolate sparse datasets, and optimised by spatial-temporal slicing to focus on desired regions. 
\item {\bf Analyse}: is the foreign interface that can be used by external systems, either via API-driven endpoints (e.g., webhooks), query interfaces, or AI-driven language models.
\item {\bf Publish}: is the long-term storage of reproducible results (e.g., for scientific publications).  It is also useful to provide online notebooks to give non-expert users a rapid development environment with all the data, code and tools.
\end{itemize}

\subsection{Agency}

The pipeline is not useful unless it meets the need of not just those in large institutions, but also is available to those who will be impacted by any policy decisions. Therefore any system for planetary computing must meet the following:
\begin{itemize}
\item {\bf Extensibility:} Can the user (rather than a platform maintainer) add new functionality to the system by importing libraries or tools to cover new techniques? Can the user easily incorporate the system into existing workflows? 
\item {\bf Accessibility:} Can it be accessed by users without asking for-- and thus possibly being denied-- permission? Can it be used by anyone without specialised computer systems knowledge? 
Does it support the patterns most useful for scientists, such as {\em incremental} data exploration and exploratory research?
\item {\bf Secrecy:} Can published results be shared in a restrictive fashion where data sensitivity is concerned, including onward sharing?
\end{itemize}

Consideration of agency is particularly important in the context of planetary computing given that many conservation actions happen in the global south, whereas much development of the computer systems involved occurs in the global north~\cite{kwet2019digital}. This requires planetary computing to take seriously the question of fairness, bias and accountability across the various computational methods, such as: how data is collected for machine learning models and how they are trained and deployed fairly~\cite{improved-fairness-ml, fairness-ml}, what role automated decision-making plays in any policy decision~\cite{automated-decision-making} and the ethical considerations for the accessibility and use of remote sensing~\cite{ethics-remote-sensing}.

\subsection{Survivability}

The system must also allow its results to survive durably as it will feed science and policy well into  the future.
\begin{itemize}
\item {\bf Traceability:} Results must be traceable through to their data and code inputs using cryptographic techniques, whilst respecting not all inputs may be directly revealed, either because of governance constraints or sensitivity (e.g., species under threat are poaching targets~\cite{yang_two_2015}). Constraints must be tracked across intermediate datasets, enforcing privacy requirements in outputs~\cite{dwork_algorithmic_2014}.
\item {\bf Explainability:}
In order to develop real-world policy based on computational results, the results and algorithms that led to them need to be understandable to a set of non-expert policymakers. This requires clear and concise expression of algorithms and introspection of ML models.
\item {\bf Reproducibility:}
The results that come out of the system need to be independently rerun, which is surprisingly difficult with heterogenous workloads spanning CPU/GPU operations~\cite{whitehead_precision_2011} and libraries varying internal algorithms across releases.
\end{itemize}

\subsection{Current state of the art}
\label{s:gap}

\begin{table*}[htb]
  \centering
  \footnotesize
  \begin{tabular}{c|l|cccc|ccc|ccc}
    & & \multicolumn{4}{c|}{End-to-end Capabilities} & \multicolumn{3}{c|}{Agency}  & \multicolumn{3}{c}{Survivability} \\
\
    & \bf Description     & \bf Ingest & \bf Transform & \bf Analyse & \bf Publish & \bf Extensible & \bf Accessible & \bf Secrecy & \bf Traceable & \bf Explainable  & \bf{Reproducible} \\
    \hline
    \parbox[t]{2mm}{\multirow{11}{*}{\rotatebox[origin=c]{90}{\bf Cloud}}}
    & Earth Engine~\cite{gorelick_google_2017}        & \chk & \chk & \chk & \chk & \crs & \chk & \chk & \crs & \crs & \crs \\
    & MPC~\cite{mpc}                                  & \chk & \chk & \chk & \chk & \chk & \crs & \chk & \crs & \chk & \chk \\
    & ArcGIS~\cite{arcgis}                            & \chk & \crs & \chk & \crs & \crs & \chk & \naa & \crs & \crs & \crs \\
    & AWS~\cite{aws}                                  & \chk & \chk & \crs & \crs & \chk & \crs & \naa & \naa & \chk & \naa \\
    & Snowflake~\cite{snowflake}                      & \crs & \chk & \crs & \crs & \chk & \crs & \naa & \naa & \chk & \naa \\
    & Huggingface~\cite{huggingface}                  & \crs & \chk & \chk & \crs & \crs & \chk & \naa & \naa & \crs & \crs \\
    & Zenodo~\cite{zenodo}                            & \naa & \naa & \naa & \chk & \crs & \chk & \crs & \chk & \chk & \naa \\
    & DataDryad~\cite{dryad}                          & \naa & \naa & \naa & \chk & \crs & \chk & \crs & \chk & \chk & \naa \\
    & GitHub~\cite{github}                            & \naa & \crs & \naa & \chk & \crs & \crs & \chk & \chk & \chk & \naa \\
    & OneDrive~\cite{onedrive}                        & \naa & \naa & \naa & \chk & \crs & \crs & \chk & \crs & \naa & \naa \\
    & Google Drive~\cite{googledrive}                 & \naa & \naa & \naa & \chk & \crs & \crs & \chk & \crs & \naa & \naa \\
    \hline
    \parbox[t]{2mm}{\multirow{6}{*}{\rotatebox[origin=c]{90}{\bf Framework}}}
    & Apache Sedona~\cite{moussa_scalable_2021}       & \crs & \chk & \chk & \crs & \crs & \chk & \naa & \crs & \chk & \chk \\
    & Apache Hadoop~\cite{hadoop}                     & \crs & \chk & \chk & \crs & \crs & \crs & \naa & \crs & \chk & \chk \\
    & Pachyderm~\cite{novella_container-based_2019}   & \crs & \chk & \chk & \crs & \crs & \crs & \naa & \crs & \chk & \chk \\
    & PostGIS~\cite{postgis}                          & \naa & \crs & \chk & \crs & \chk & \crs & \naa & \crs & \chk & \chk \\
    & TileDB~\cite{tiledb}                            & \naa & \chk & \chk & \crs & \crs & \crs & \naa & \crs & \chk & \chk \\
    & Jenkins~\cite{jenkins}                          & \naa & \crs & \naa & \chk & \chk & \crs & \crs & \chk & \chk & \chk \\
    \hline
    \parbox[t]{2mm}{\multirow{5}{*}{\rotatebox[origin=c]{90}{\bf Libraries}}}
    & GDAL~\cite{gdal}                                & \naa & \crs & \crs  & \naa & \crs & \chk & \naa & \naa & \chk & \chk \\
    & CUPY~\cite{cupy}                                & \naa & \crs & \crs  & \naa & \crs & \crs & \naa & \naa & \chk & \chk \\
    & Dask~\cite{dask}                                & \naa & \chk & \crs  & \naa & \chk & \chk & \naa & \naa & \naa & \chk \\
    & TerraLib~\cite{terralib}                        & \naa & \chk & \chk  & \naa & \crs & \crs & \naa & \naa & \chk & \chk \\
    & PyTorch~\cite{paszke_pytorch_2019}              & \naa & \chk & \chk  & \naa & \crs & \chk & \naa & \naa & \crs & \chk \\
    \hline
    \parbox[t]{2mm}{\multirow{3}{*}{\rotatebox[origin=c]{90}{\bf Tools}}}
    & R~\cite{rstats}                                 & \chk & \crs & \chk & \naa & \chk & \chk & \naa & \crs & \chk & \chk\\
    & QGIS~\cite{qgis}                                & \naa & \crs & \chk & \naa & \crs & \chk & \naa & \naa & \naa & \naa \\
    & STAC~\cite{stac}                                & \chk & \naa & \naa & \chk & \chk & \chk & \naa & \crs & \chk & \crs \\
    \hline
  \end{tabular}
  \caption{\label{t:gaap} Comparison of existing tools used for environmental science {\em (a dash indicates an out-of-scope area)}}
\end{table*}

We survey existing systems via two strategies: use existing end-to-end platforms that cover as much of the lifecycle we have identified as possible, or pull together a custom system using off-the-shelf components. Our survey is summarized in Table~\ref{t:gaap}.

\paragraph{Existing end-to-end solutions}

Google's Earth Engine (GEE)~\cite{gorelick_google_2017} and Microsoft's Planetary Computer (MPC)~\cite{mpc} both provide a cloud-based end-to-end solution targeted at ecologists. These platforms already ingest and make available popular satellite imagery, provide ways to interactively explore existing datasets, and make it easy to process the data using familiar languages such as Python and JavaScript, whilst hiding the computation complexity of scaling your algorithm. The two platforms differ in their approach, with GEE focusing on ease-of-use through its own proprietary UI and libraries, and MPC focusing on providing an open platform on which to run existing tooling at scale.
Both platforms score well on E2E capabilities, but less so on the non-technical requirements. Access to both is on a request basis with no guarantee of access, and therefore no guarantee that any work can be reproduced. Code written for GEE will not run directly on other platforms, also hampering reproducibility compared to MPC's more open stance. Conversely, MPC exposes systems concerns to its users in a far greater way than GEE, which uses its own frameworks to abstract memory and parallelism.
Neither system supports traceability and explainability while protecting private data.

\paragraph{Platforms}
The open source community has produced a number of components over the years to build custom solutions (we only cover a small sample here). These typically expose more system concerns to the user and require more management.
The size and complexity of datasets required to tackle global environmental problems is petabytes in scale and always growing, equating to large storage and processing costs.
Many turn to general-purpose cloud platforms (GPCPs) as an affordable, scalable alternative, using services like Amazon Web Services (AWS)~\cite{aws} and Snowflake~\cite{snowflake}. Some, like AWS, already have open science datasets ready to go, though that is not common. GPCPs are much less ready for non-technologists to use without additional training: configuring infrastructure necessarily exposes systems complexities. Users need to ensure their architecture matches the infrastructure's mechanisms for CPU/GPU/memory provisioning, something GEE hides well.

Alternatively, there are on-prem solutions that are suitable for spinning up a scalable compute platform on local hardware using generic frameworks like Hadoop~\cite{hadoop} or Pachyderm~\cite{novella_container-based_2019}, and geospatial targeted ones like Sedona~\cite{moussa_scalable_2021}. These frameworks schedule storage and compute resources, but do require both technical setup and maintenance. These systems will favour a particular class of algorithms: e.g., Hadoop's map-reduce makes the user structure their code to align with that, and this doesn't work efficiently for algorithms such as deep learning~\cite{murray_ciel_2011}.
There are a variety of GIS enabled databases (such as PostGIS~\cite{postgis} and TileDB~\cite{tiledb}) that make working with certain classes of geo-data accessible, which are powerful tools once set up, but knowing how to run a high-performance database with very large volumes of data is a job in its own right. 

For incrementality, Continuous Integration (CI) systems (such as Jenkins~\cite{jenkins}) can be set up to help with running end-to-end pipelines, but are remarkably difficult to make reproducible and portable across systems, and again are typically managed by dedicated devops professionals due to the their fragility and lack of end-user friendliness. GitHub~\cite{github} does provide some easily managed CI style features that are easier to set up, but are not intended to be used for computation at a planetary scale.

For publishing, hosted services for code and data (e.g., GitHub, Zenodo~\cite{zenodo} and DataDryad~\cite{dryad}) provide a degree of traceable public reference by both being a primary source and allowing for versioning. However, ensuring that private data is never leaked across these services is difficult as DIFC has not been adopted~\cite{zeldovich_securing_2008}. It is notable that despite the adaption by academics of data portals like these, major institutions like NASA and the JRC still self-publish, leading to issues like those described in~\S\ref{section:data}. 
Typically these platforms are only used for final published results, but in practice ecologists regularly need to share work-in-progress results with colleagues in other institutions, yet geospatial datasets will readily exceed the size limits of most people's cloud provider file stores (e.g., Google Drive or Microsoft OneDrive~\cite{googledrive, onedrive}), and so ad-hoc solutions are used, which limit traceability.

\paragraph{Data transform}
Coding efficiently at the scales involved with geospatial datasets often requires CS-level knowledge to use hardware like GPUs. Conventional tooling used in this domain, e.g., Python or R, do not sufficiently abstract memory management and task parallelism to allow the user to express their transformation and analysis algorithms efficiently alone, but there exist layered frameworks that can help. For example, PyTorch~\cite{paszke_pytorch_2019} provides a ``pragmatic'' programming interface to simplify deep learning by abstracting parallelism and GPU use. Dask~\cite{dask} does similar for NumPy-based numeric workloads, distributing them over multiple cores or over a cluster. This ability to separate the expression of the computation to the management of the data resources is a key enabler of users that need to be able to work on these datasets. 

Frameworks only provide a partial solution to the data lifecycle and the boundaries expose systems-level concerns. For instance, CuPy~\cite{cupy}, a CUDA-based GPU based drop-in replacement for the numerical framework NumPy~\cite{harris_array_2020}, does not consider whether it is efficient to use the GPU over the CPU for a particular dataset: it leaves the burden of that with the user, assuming they are familiar with details like having to do enough work on the GPU to amortise memory transfer costs. Interactions between frameworks can also cause significant performance issues: GDAL~\cite{gdal}, the standard library for reading and writing geo-data, typically suggests an efficient block-size for reading GeoTIFFs that is orientated incompatibly with CuPy's expectations, leading to just one ALU on the GPU being used unless the user knows to redimension their data first. In terms of uncertainty, whilst the ability to migrate data between CPU and GPU is convenient, it also leads to varying results depending on the sensitivity of the experiment, as the floating point behaviour of GPUs differs from those in CPUs~\cite{whitehead_precision_2011}.

\paragraph{Analysis and Visualisation} 
By using common data-format standards in the domain (e.g., GeoJSON, GeoTIFFs etc.), interoperating with existing analysis tools such as Leaflet~\cite{leaflet} and QGIS~\cite{qgis} for exploratory visualisation and R~\cite{rstats} for statistical analysis becomes much easier. These are crucial in ecology and geospatial analysis for enabling visual and statistical debugging. For example, plotting points on a map makes it much easier to know when your latitudes and longitudes are stored the wrong way round. Integrating these tools more deeply into computational pipelines can help improve the process of checking the progress and correctness of a calculation. 
Being able to visualise datasets is so important that we have observed cases where when given a tradeoff between an optimal geospatial data format (e.g., not using raster data due to the limitations of map projections and instead using uniform-area hexgrids), ecologists fall back to using formats they can visualise.

\paragraph{Publication and traceability}

Despite data lineage in science long being a concern~\cite{simmhan_survey_2005}, common geospatial data-storage formats do not provide support for versioning, let alone traceability. GeoTIFF, the common format for satellite imagery data, contains no fields around versioning in the standard~\cite{devys_geotiff_2019}, and nor do upcoming formats like geozarr~\cite{GeoZarr}. The Arrow and Parquet formats have provision for embedded metadata to use for data lineage, but there are no mandatory fields or common conventions.
SpatioTemporal Asset Catalogs (STAC)~\cite{stac} is a JSON-based format for publishing geo-data collections, and is used by many services including MPC. STAC has extensions that capture how data was processed (which helps with reproducability) but has no mandatory versioning information for the referenced data-files, and the optional version info does not contain anything (e.g., data hashes) to ensure versions are tied to specific data instances.

Current end-to-end solutions are let down by lack of reproducibility, traceability, and explainability -- all key to ensuring the right decisions are made when tackling the climate crisis.
Whilst it is possible to assemble off-the-shelf components into an end-to-end system with these features, in doing so they expose the user to system concerns that require a technology domain expert, acting as a blocker to wider adaptation of the environmental interventions that they would otherwise enable.

\section{Open research challenges}
\label{s:discuss}
\label{sec:approach}

\begin{figure}
    \centering
    \includegraphics[width=0.99\linewidth]{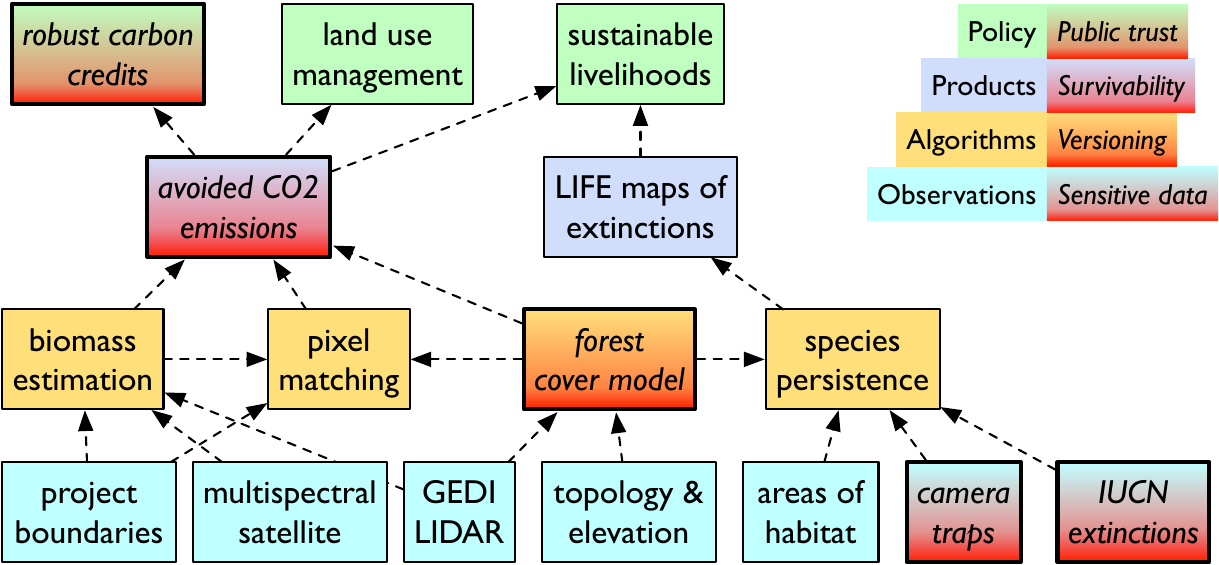}
    \caption{The pipeline from Figure \ref{fig:pos-before}, but highlighting computing research challenges around privacy, versioning and longevity. The computing and environmental challenges depend on each other for a trustworthy policy outcome.}
    \label{fig:pos-after}
    \vspace{-1em}
\end{figure}

We have established there are missing pieces (\S\ref{s:requirements}) needed to reduce the difficulty of implementing data-driven environmental policy-making  (\S\ref{s:motive}), and that no single computing framework or cloud platform solves it all. The heart of the problem is the sheer amount of uncertainty introduced by the layers of global observations, algorithms, derived products and eventual policy-making (\S\ref{s:lessons}).
We now explore computer science research challenges which, if connected, pave the way for a sustainable planetary computing model. Figure~\ref{fig:pos-after} summarises the general opportunities we discovered while building our own digital pipelines. We must deal with sensitive data observations (\S\ref{sec:privacy}), then how to version these large environments (\S\ref{sec:version}), how to ensure they will survive for decades (\S\ref{sec:survive}) and finally meet our goal of building trust in data-driven policy-making (\S\ref{sec:publictrust}).

\subsection{Reconciling privacy and transparency}
\label{sec:privacy}

Climate change monitoring and environmental conservation are both hugely challenging from a data secrecy perspective, and fully open access to data is likely to cause more harm than good. There are a number of stakeholders involved in any data collection process, with numerous examples of environmental data being used by economically-motivated malicious actors to cause harm (\S\ref{section:data}).

From a computer science perspective, we thus need to build in strong security and privacy controls from the ground up in any planetary computing infrastructure.  While most existing systems prioritise openness or transparency at the expense of privacy, we have established that being too open lets bad actors determine how best to game the system~\cite{richards_are_2017}. To mitigate this, as one possibility, we suggest the principle of ``eventual openness'' where data is initially embargoed and eventually made public ~\cite{lowe_publish_2017}. Moreover, differential privacy~\cite{dwork_algorithmic_2014} and decentralised information flow control~\cite{zeldovich_securing_2008} permit some transparency while preserving data privacy even during the sensitive early period.
Full query engines that respect the privacy constraints across multiple users are also an emerging area; e.g. the multiverse database architecture~\cite{marzoev_towards_2019}. These capabilities could provide scientists and policymakers with the tools they need to balance privacy and transparency.
It may also be useful to {\em partially} reveal source data to avoid full disclosure, but allow subsequent auditing by third-parties who are granted access to the source information and can independently verify it.

\subsection{Versioning uncertainty across code and data}
\label{sec:version}

We observed (\S\ref{section:data}) that even seemingly static datasets for past observations can vary over time due to the algorithms that process them improving. Figure~\ref{fig:pos-after} shows this can impact downstream policy-making such as carbon credits, which leads to adverse effects for the desired conservation outcomes. It is therefore important that we establish usable versioning schemes {\em across} code and datasets and the environments they are processed in, in order for policy-makers to be able to predict the impact of upcoming changes.

There exist initiatives to structure data formats involved in climate modelling (e.g., for CMIP6~\cite{balaji2018requirements}). The key they identified to efficiency is to establish a common data format that avoids repeated data transformations in subsequent computation pipelines.
While there are many workflow engines available, none support portable and efficient pipeline composition due to the varying expressive power of dynamic dataflow graphs; from Turing-powerful engines~\cite{murray_ciel_2011} to static graphs that do not support data-dependent control flow~\cite{hadoop}.
A protocol-driven approach could specify library interfaces for expressing multiple code stages in different languages, orchestrating container builds, GPU execution, and results retrieval from external systems (e.g., GEE) where no local copy is available.  As with ingestion, systems must propagate secrecy information of datasets across distributed nodes~\cite{zeldovich2008securing}.
A planetary computing pipeline thus needs to account for traceability, privacy and sharing while adding versioning, and some research directions to investigate include:
\one{}
using compressed columnar formats sorted by anticipated access patterns~\cite{saeedan2022} with optional lossy compression~\cite{dicapello2016, esriLERC} optimised for streaming to CPUs/GPUs~\cite{tiancusz2020};
\two{}
tagging input sources using Decentralised Information Flow Control (DIFC) labels~\cite{zeldovich_securing_2008} ensuring that access control checks can be applied at any point in a distributed pipeline or a query engine~\cite{marzoev_towards_2019};
\three{}
hashing the data into spatial chunks, permitting subsequent version control~\cite{gazagnaire_irminsule_2014} of subsets.

\subsection{Building survivable data products}
\label{sec:survive}

At first glance, a cloud platform like GEE solves the short-term planetary computing problem (Table~\ref{t:gaap}) and is being used widely~\cite{mutanga2019google}. The challenge is how to make it survive into the coming decades. It is difficult to believe that a closed platform -- no matter how impressively engineered -- can gather the world's data and act as a central source of policy-making intelligence.
The computing community has come together in the past to build {\em federated} testbeds for emerging technologies, such as PlanetLab~\cite{peterson_design_2006}, CloudLab~\cite{duplyakin2019design} and Emulab~\cite{hermenier_how_2012}.  Since the inception of the Internet, we have built collective knowledge about fostering and scaling open source communities~\cite{tan2020scaling}, curating open data collections~\cite{parsons_unlocking_2023}, and drive shared governance such as the Internet Engineering Task Force (IETF)~\cite{harcourt2020internal}.

Planetary computing needs a similarly ambitious drive to ensure that platforms such as GEE can continue to survive beyond any one organisation or provider. It is not yet clear if the answer to this is by building a fully federated system, but a first step in that direction is to 
define {\em protocols} by which independent implementations can interoperate. An IETF-equivalent process would involve specifying protocols for each of the pipeline blocks in Figure~\ref{f:ark}. However, there are significant gaps in the research literature with respect to decentralised access control and encryption at scale.

\begin{figure}
    \centering
    \includegraphics[width=0.99\linewidth]{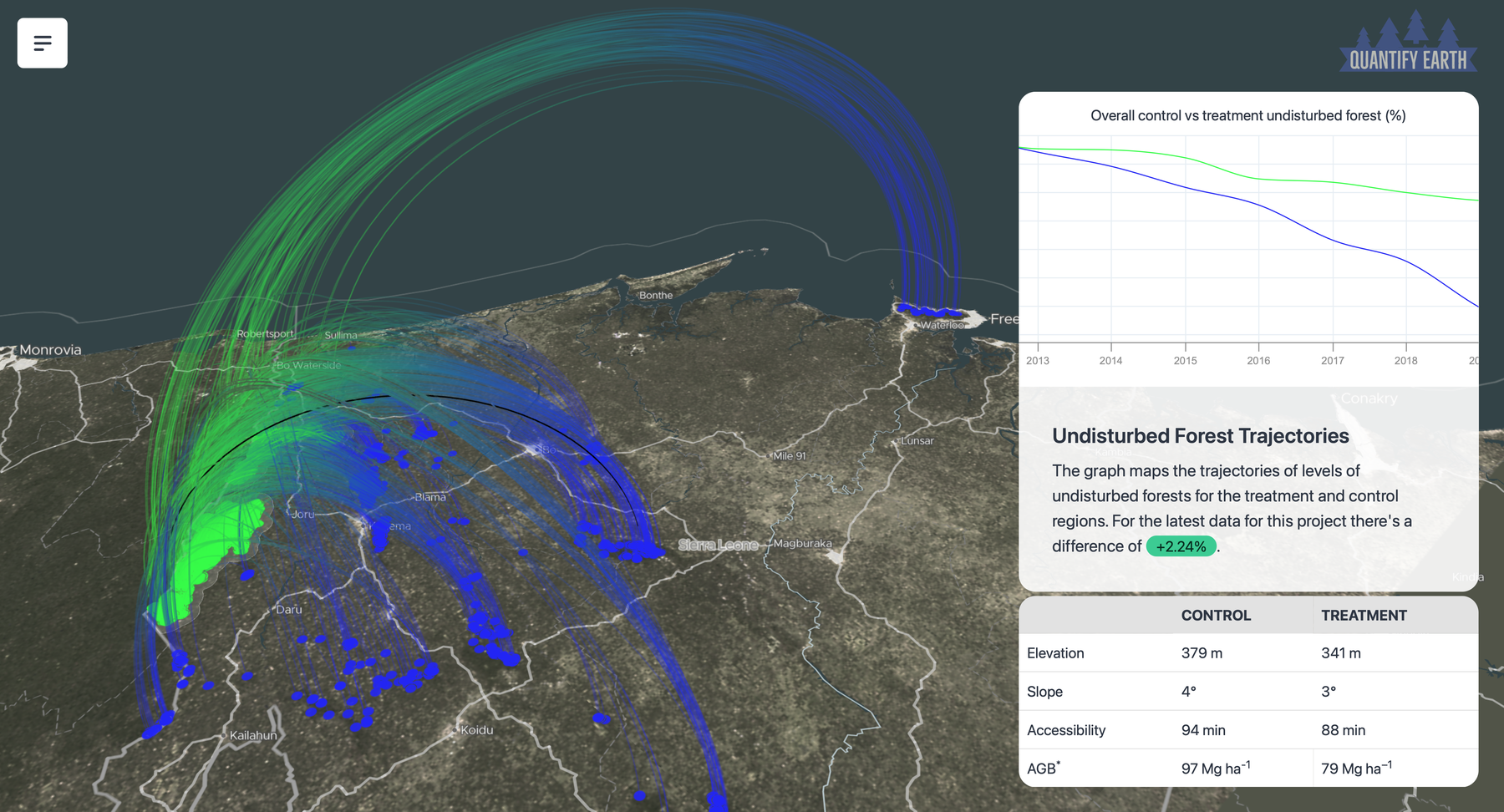}
    \caption{A visualisation of the counterfactual pixel-matching to assess the additionality of an avoided deforestation project (\S\ref{s:motive}). Pixels in green are from the project and they are matched to similar locations in blue.}
    \label{fig:qe}
    \vspace{-1em}
\end{figure}

\subsection{Accessibility to non-expert users}
\label{sec:publictrust}

Scientists and policymakers are the \textit{raison d'\^{e}tre} and primary users of planetary computing, not an afterthought, and computing must be deployed towards positive terminal outcomes~\cite{seth2022technology} to those affected most by climate change. The purpose of the data pipelines (Figure~\ref{fig:pos-after}) is to empower us to analyse complex counterfactual scenarios in order to enact difficult policy decisions. Therefore, we need simpler query interfaces to the lower level expert-driven data products, while preserving the provenance and explainability of those products to retain trust in the resulting decisions.
In our own research while working on tropical forests and carbon credits~\cite{cowg}, one of the most impactful interfaces was a visual explorer which illustrates the counterfactual analysis involved step-by-step (Figure~\ref{fig:qe}). This dynamic interface began as a prototype to debug the algorithms, but rapidly turned into the explainer of the complex algorithms~\cite{tmfv2} to ecologists and financiers.  Promising research directions include ``no-code'' interfaces to assemble building blocks of algorithms and data~\cite{cai2023lowcode}, and LLM-based natural language copilots for conversational exploration resulting in reusable outputs~\cite{busch2023chatgpt}.

The other major consideration is that of agency and independence -- it must be possible for policymakers to mix public datasets with their own private (usually jurisdiction-specific) data without revealing it to a third party~\cite{rspet}. The safe mixing of such data requires careful analysis to establish biases that might adversely affect its applicability to a given region. Approaches such as datasheets for datasets~\cite{gebru2021datasheets} may help establish standards for helping non-expert users ensure the sources they are basing their decisions on are appropriate for their local context.

\section{Conclusions}

We have drawn on our experiences as a joint team of computing and ecological scientists to lay out our vision for planetary computing as a vehicle to accelerating collaborative, data-driven environmental policy-making.
The research directions we have outlined here are by no means comprehensive, and are primarily focused on computer systems research. We hope, however, that the framing of planetary computing in this paper will act as a useful focal point for the computer science community to plug key knowledge gaps in the deployment of data-driven environmental science, and will eventually result in an accurate and representative view of planetary health that can be used to inform policy that preserves and enhances our natural ecosystems into the coming decades.

{\footnotesize
\bibliographystyle{acm}
\bibliography{pos,cucl-eeg}

\begin{thebibliography}{100}

\bibitem{postgis}
About {PostGIS}.
\newblock \url{https://postgis.net/}.

\bibitem{zenodo}
About zenodo.
\newblock \url{https://about.zenodo.org/}.

\bibitem{hadoop}
{Apache} {Hadoop}.
\newblock \url{https://hadoop.apache.org}.

\bibitem{arcgis}
{ArcGIS} {Online}.
\newblock
  \url{https://www.esri.com/en-us/arcgis/products/arcgis-online/overview}.

\bibitem{cupy}
{CuPy}: {NumPy} and {SciPy} for {GPU}.
\newblock \url{https://cupy.dev/}.

\bibitem{dask}
Dask | scale the python tools you love.
\newblock \url{https://dask.org/}.

\bibitem{dryad}
Dryad - publish and preserve your data.
\newblock \url{https://datadryad.org/}.

\bibitem{gdal}
{GDAL}.
\newblock \url{https://gdal.org/}.

\bibitem{GeoZarr}
Geozarr-spec.
\newblock \url{https://github.com/zarr-developers/geozarr-spec}.

\bibitem{github}
{GitHub}.
\newblock \url{https://github.com/}.

\bibitem{huggingface}
{HuggingFace}.
\newblock \url{https://huggingface.co/}.

\bibitem{jenkins}
Jenkins.
\newblock \url{https://www.jenkins.io/}.

\bibitem{esriLERC}
{LERC} - {Limited} {Error} {Raster} {Compression}.
\newblock \url{https://github.com/Esri/lerc}.

\bibitem{mpc}
{Microsoft Planetary Computer}.
\newblock \url{https://planetarycomputer.microsoft.com/}.

\bibitem{aws}
{Overview} of {Amazon} {Web} {Services}.
\newblock
  \url{https://docs.aws.amazon.com/whitepapers/latest/aws-overview/introduction.html}.

\bibitem{googledrive}
Personal cloud storage and file sharing platform - google.
\newblock \url{https://www.google.com/intl/en-US/drive/}.

\bibitem{onedrive}
Personal cloud storage – microsoft onedrive.
\newblock
  \url{https://www.microsoft.com/en-gb/microsoft-365/onedrive/online-cloud-storage}.

\bibitem{snowflake}
Snowflake.
\newblock \url{https://www.snowflake.com/}.

\bibitem{stac}
{STAC}: {SpatioTemporal} {Asset} {Catalogs}.
\newblock \url{https://stacspec.org/}.

\bibitem{terralib}
{TerraLib5}.
\newblock \url{http://www.terralib.org/}.

\bibitem{tiledb}
{TileDB} - {The} {Universal} {Database}.
\newblock \url{https://tiledb.com/}.

\bibitem{leaflet}
{Leaflet}: an open-source {JavaScript} library for mobile-friendly interactive
  maps.
\newblock \url{https://leafletjs.com}, 2023.

\bibitem{jrc-update}
List of updates integrated in the {TMF} 2022 products.
\newblock \url{https://forobs.jrc.ec.europa.eu/TMF/data\#update}, 2023.

\bibitem{qgis}
{QGIS} - {A} {Free} and {Open} {Source} {Geographic} {Information} {System}.
\newblock \url{https://www.qgis.org/}, 2023.

\bibitem{rstats}
The r project for statistical computing.
\newblock \url{https://www.r-project.org}, 2023.

\bibitem{achard2009monitoring}
{\sc Achard, F., Beuchle, R., Bodart, C., Brink, A., Carboni, S., Eva, H.,
  Mayaux, P., Ra{\v{s}}i, R., Simonetti, D., and Stibig, H.-J.}
\newblock Monitoring forest cover at global scale: The jrc approach.
\newblock In {\em Proceedings of the 33rd ISRSE conference\/} (2009), vol.~4.

\bibitem{angelsen2009realising}
{\sc Angelsen, A.}
\newblock {\em Realising REDD+: National strategy and policy options}.
\newblock Cifor, 2009.

\bibitem{balaji2018requirements}
{\sc Balaji, V., Taylor, K.~E., Juckes, M., Lawrence, B.~N., Durack, P.~J.,
  Lautenschlager, M., Blanton, C., Cinquini, L., Denvil, S., Elkington, M.,
  et~al.}
\newblock Requirements for a global data infrastructure in support of cmip6.
\newblock {\em Geoscientific Model Development 11}, 9 (2018), 3659--3680.

\bibitem{doi:10.1126/science.adh3426}
{\sc Balmford, A., Brancalion, P. H.~S., Coomes, D., Filewod, B., Groom, B.,
  Guizar-Coutiño, A., Jones, J. P.~G., Keshav, S., Kontoleon, A.,
  Madhavapeddy, A., Malhi, Y., Sills, E.~O., Strassburg, B. B.~N., Venmans, F.,
  West, T. A.~P., Wheeler, C., and Swinfield, T.}
\newblock Credit credibility threatens forests.
\newblock {\em Science 380}, 6644 (2023), 466--467.

\bibitem{tmfv2}
{\sc Balmford, A., Coomes, D., Hartup, J., Jaffer, S., Keshav, S., Lam, M., and
  Wheeler, C.}
\newblock Pact tropical moist forest accreditation methodology.
\newblock Tech. rep., 2023.

\bibitem{balmford_how_2019}
{\sc Balmford, B., Green, R.~E., Onial, M., Phalan, B., and Balmford, A.}
\newblock How imperfect can land sparing be before land sharing is more
  favourable for wild species?
\newblock {\em Journal of Applied Ecology 56}, 1 (2019), 73--84.
\newblock \_eprint:
  https://onlinelibrary.wiley.com/doi/pdf/10.1111/1365-2664.13282.

\bibitem{barnes_publish_2010}
{\sc Barnes, N.}
\newblock Publish your computer code: it is good enough.
\newblock {\em Nature 467}, 7317 (Oct. 2010), 753--753.

\bibitem{9833110}
{\sc Blinn, A., Moon, D., Griffis, E., and Omar, C.}
\newblock An integrative human-centered architecture for interactive
  programming assistants.
\newblock In {\em 2022 IEEE Symposium on Visual Languages and Human-Centric
  Computing (VL/HCC)\/} (2022), pp.~1--5.

\bibitem{brooks2019measuring}
{\sc Brooks, T.~M., Pimm, S.~L., Ak{\c{c}}akaya, H.~R., Buchanan, G.~M.,
  Butchart, S.~H., Foden, W., Hilton-Taylor, C., Hoffmann, M., Jenkins, C.~N.,
  Joppa, L., et~al.}
\newblock Measuring terrestrial area of habitat (aoh) and its utility for the
  iucn red list.
\newblock {\em Trends in ecology \& evolution 34}, 11 (2019), 977--986.

\bibitem{buchwitz2018copernicus}
{\sc Buchwitz, M., Reuter, M., Schneising, O., Bovensmann, H., Burrows, J.~P.,
  Boesch, H., Anand, J., Parker, R., Detmers, R.~G., Aben, I., et~al.}
\newblock Copernicus climate change service (c3s) global satellite observations
  of atmospheric carbon dioxide and methane.
\newblock {\em Advances in Astronautics Science and Technology 1\/} (2018),
  57--60.

\bibitem{burgess_harnessing_2010}
{\sc Burgess, S., Kranz, M., Turner, N., Cardell-Oliver, R., and Dawson, T.}
\newblock Harnessing wireless sensor technologies to advance forest ecology and
  agricultural research.
\newblock {\em Agricultural and Forest Meteorology 150}, 1 (Jan. 2010), 30--37.

\bibitem{busch2023chatgpt}
{\sc Busch, D., Nolte, G., Bainczyk, A., and Steffen, B.}
\newblock Chatgpt in the loop: a natural language extension for domain-specific
  modeling languages.
\newblock In {\em International Conference on Bridging the Gap between AI and
  Reality\/} (2023), Springer, pp.~375--390.

\bibitem{cai2023lowcode}
{\sc Cai, Y., Mao, S., Wu, W., Wang, Z., Liang, Y., Ge, T., Wu, C., You, W.,
  Song, T., Xia, Y., Tien, J., and Duan, N.}
\newblock Low-code llm: Visual programming over llms, 2023.

\bibitem{animal-tracking}
{\sc Cooke, S.~J., Nguyen, V.~M., Kessel, S.~T., Hussey, N.~E., Young, N., and
  Ford, A.~T.}
\newblock Troubling issues at the frontier of animal tracking for conservation
  and management.
\newblock {\em Conservation Biology 31}, 5 (2017), 1205--1207.

\bibitem{lowavail}
{\sc Culina, A., van~den Berg, I., Evans, S., and Sánchez-Tójar, A.}
\newblock Low availability of code in ecology: A call for urgent action.
\newblock {\em PLOS Biology 18}, 7 (07 2020), 1--9.

\bibitem{devys_geotiff_2019}
{\sc Devys, E., Habermann, T., Heazel, C., Lott, R., and Rouault, E.}
\newblock {OGC} {GeoTIFF} {Standard}.

\bibitem{dicapello2016}
{\sc Di, S., and Cappello, F.}
\newblock Fast error-bounded lossy hpc data compression with sz.
\newblock In {\em 2016 IEEE International Parallel and Distributed Processing
  Symposium (IPDPS)\/} (2016), pp.~730--739.

\bibitem{nix}
{\sc Dolstra, E., and de~Jonge, M.}
\newblock Nix: A safe and {Policy-Free} system for software deployment.
\newblock In {\em 18th Large Installation System Administration Conference
  (LISA 04)\/} (Atlanta, GA, Nov. 2004), USENIX Association.

\bibitem{dubayah2022gedi}
{\sc Dubayah, R., Armston, J., Healey, S.~P., Bruening, J.~M., Patterson,
  P.~L., Kellner, J.~R., Duncanson, L., Saarela, S., St{\aa}hl, G., Yang, Z.,
  et~al.}
\newblock Gedi launches a new era of biomass inference from space.
\newblock {\em Environmental Research Letters 17}, 9 (2022), 095001.

\bibitem{dubayah2020global}
{\sc Dubayah, R., Blair, J.~B., Goetz, S., Fatoyinbo, L., Hansen, M., Healey,
  S., Hofton, M., Hurtt, G., Kellner, J., Luthcke, S., et~al.}
\newblock The global ecosystem dynamics investigation: High-resolution laser
  ranging of the earth’s forests and topography.
\newblock {\em Science of remote sensing 1\/} (2020), 100002.

\bibitem{duplyakin2019design}
{\sc Duplyakin, D., Ricci, R., Maricq, A., Wong, G., Duerig, J., Eide, E.,
  Stoller, L., Hibler, M., Johnson, D., Webb, K., et~al.}
\newblock The design and operation of cloudlab.
\newblock In {\em 2019 USENIX annual technical conference (USENIX ATC 19)\/}
  (2019), pp.~1--14.

\bibitem{dwork_algorithmic_2014}
{\sc Dwork, C., and Roth, A.}
\newblock {\em The {Algorithmic} {Foundations} of {Differential} {Privacy}}.
\newblock Now Foundations and Trends, 2014.

\bibitem{eyres_life_2023}
{\sc Eyres, A., Ball, T., Dales, M., Swinfield, T., Arnell, A., Baisero, D.,
  Durán, A.~P., Green, J., Madhavapeddy, A., and Balmford, A.}
\newblock {LIFE}: {A} metric for quantitively mapping the impact of land-cover
  change on global extinctions, Nov. 2023.

\bibitem{freckleton_census_2006}
{\sc Freckleton, R.~P., Watkinson, A.~R., Green, R.~E., and Sutherland, W.~J.}
\newblock Census error and the detection of density dependence.
\newblock {\em The Journal of Animal Ecology 75}, 4 (July 2006), 837--851.

\bibitem{gazagnaire_irminsule_2014}
{\sc Gazagnaire, T., Chaudhry, A., Madhavapeddy, A., Mortier, R., Scott, D.,
  Sheets, D., Tsipenyuk, G., and Crowcroft, J.}
\newblock Irminsule: a branch-consistent distributed library database.
\newblock In {\em 4th {ACM} {OCaml} {Users} and {Developers} {Workshop}\/}
  (Gothenbergh, Sweden, Sept. 2014), ACM.

\bibitem{gebru2021datasheets}
{\sc Gebru, T., Morgenstern, J., Vecchione, B., Vaughan, J.~W., Wallach, H.,
  Iii, H.~D., and Crawford, K.}
\newblock Datasheets for datasets.
\newblock {\em Communications of the ACM 64}, 12 (2021), 86--92.

\bibitem{stadia}
{\sc Gerken, T.}
\newblock Gamers say goodbye to google's stadia as platform shuts, 2023.
\newblock Accessed on October 23, 2023.

\bibitem{gorelick_google_2017}
{\sc Gorelick, N., Hancher, M., Dixon, M., Ilyushchenko, S., Thau, D., and
  Moore, R.}
\newblock Google {Earth} {Engine}: {Planetary}-scale geospatial analysis for
  everyone.
\newblock {\em Remote Sensing of Environment 202\/} (Dec. 2017), 18--27.

\bibitem{gee}
{\sc Gorelick, N., Hancher, M., Dixon, M., Ilyushchenko, S., Thau, D., and
  Moore, R.}
\newblock Google earth engine: Planetary-scale geospatial analysis for
  everyone. re-mote sensing of environment, 202, 18--27, 2017.

\bibitem{grover2001one}
{\sc Grover, J.}
\newblock One cast beyond--the public’s right to know--radiotelemetry.
\newblock {\em Fisherman 26}, 5 (2001), 18--22.

\bibitem{harcourt2020internal}
{\sc Harcourt, A., Christou, G., and Simpson, S.}
\newblock Internal governance of the ietf, w3c and ieee: structure,
  decision-making and internationalisation.
\newblock In {\em Global Standard Setting in Internet Governance}. Oxford
  University Press, 2020.

\bibitem{harfoot_using_2021}
{\sc Harfoot, M. B.~J., Johnston, A., Balmford, A., Burgess, N.~D., Butchart,
  S. H.~M., Dias, M.~P., Hazin, C., Hilton-Taylor, C., Hoffmann, M., Isaac, N.
  J.~B., Iversen, L.~L., Outhwaite, C.~L., Visconti, P., and Geldmann, J.}
\newblock Using the {IUCN} {Red} {List} to map threats to terrestrial
  vertebrates at global scale.
\newblock {\em Nature Ecology \& Evolution 5}, 11 (Nov. 2021), 1510--1519.
\newblock Number: 11 Publisher: Nature Publishing Group.

\bibitem{harris_array_2020}
{\sc Harris, C.~R., Millman, K.~J., van~der Walt, S.~J., Gommers, R., Virtanen,
  P., Cournapeau, D., Wieser, E., Taylor, J., Berg, S., Smith, N.~J., Kern, R.,
  Picus, M., Hoyer, S., van Kerkwijk, M.~H., Brett, M., Haldane, A., del Río,
  J.~F., Wiebe, M., Peterson, P., Gérard-Marchant, P., Sheppard, K., Reddy,
  T., Weckesser, W., Abbasi, H., Gohlke, C., and Oliphant, T.~E.}
\newblock Array programming with {NumPy}.
\newblock {\em Nature 585}, 7825 (Sept. 2020), 357--362.
\newblock Number: 7825 Publisher: Nature Publishing Group.

\bibitem{hermenier_how_2012}
{\sc Hermenier, F., and Ricci, R.}
\newblock How to {Build} a {Better} {Testbed}: {Lessons} from a {Decade} of
  {Network} {Experiments} on {Emulab}.
\newblock In {\em Testbeds and {Research} {Infrastructure}. {Development} of
  {Networks} and {Communities}\/} (Berlin, Heidelberg, 2012), T.~Korakis,
  M.~Zink, and M.~Ott, Eds., Lecture {Notes} of the {Institute} for {Computer}
  {Sciences}, {Social} {Informatics} and {Telecommunications} {Engineering},
  Springer, pp.~287--304.

\bibitem{herndon-spreadsheet}
{\sc Herndon, T., Ash, M., and Pollin, R.}
\newblock {Does high public debt consistently stifle economic growth? A
  critique of Reinhart and Rogoff}.
\newblock {\em Cambridge Journal of Economics 38}, 2 (12 2013), 257--279.

\bibitem{HOLCOMB2023100106}
{\sc Holcomb, A., Mathis, S.~V., Coomes, D.~A., and Keshav, S.}
\newblock Computational tools for assessing forest recovery with gedi shots and
  forest change maps.
\newblock {\em Science of Remote Sensing 8\/} (2023), 100106.

\bibitem{improved-fairness-ml}
{\sc Holstein, K., Wortman~Vaughan, J., Daum\'{e}, H., Dudik, M., and Wallach,
  H.}
\newblock Improving fairness in machine learning systems: What do industry
  practitioners need?
\newblock In {\em Proceedings of the 2019 CHI Conference on Human Factors in
  Computing Systems\/} (New York, NY, USA, 2019), CHI '19, Association for
  Computing Machinery, p.~1–16.

\bibitem{hu_northward_2022}
{\sc Hu, Q., and Han, Z.}
\newblock Northward {Expansion} of {Desert} {Climate} in {Central} {Asia} in
  {Recent} {Decades}.
\newblock {\em Geophysical Research Letters 49}, 11 (2022), e2022GL098895.
\newblock \_eprint:
  https://onlinelibrary.wiley.com/doi/pdf/10.1029/2022GL098895.

\bibitem{jakob_need_2023}
{\sc Jakob, C., Gettelman, A., and Pitman, A.}
\newblock The need to operationalize climate modelling.
\newblock {\em Nature Climate Change 13}, 11 (Nov. 2023), 1158--1160.
\newblock Number: 11 Publisher: Nature Publishing Group.

\bibitem{justice2002overview}
{\sc Justice, C., Townshend, J., Vermote, E., Masuoka, E., Wolfe, R., Saleous,
  N., Roy, D., and Morisette, J.}
\newblock An overview of modis land data processing and product status.
\newblock {\em Remote sensing of Environment 83}, 1-2 (2002), 3--15.

\bibitem{mashup-vcs}
{\sc Kuttal, S.~K., Sarma, A., and Rothermel, G.}
\newblock History repeats itself more easily when you log it: Versioning for
  mashups.
\newblock In {\em 2011 IEEE Symposium on Visual Languages and Human-Centric
  Computing (VL/HCC)\/} (2011), pp.~69--72.

\bibitem{kwet2019digital}
{\sc Kwet, M.}
\newblock Digital colonialism: Us empire and the new imperialism in the global
  south.
\newblock {\em Race \& Class 60}, 4 (2019), 3--26.

\bibitem{lindenmayer_not_2017}
{\sc Lindenmayer, D., and Scheele, B.}
\newblock Do not publish.
\newblock {\em Science 356}, 6340 (May 2017), 800--801.
\newblock Publisher: American Association for the Advancement of Science.

\bibitem{lowe_publish_2017}
{\sc Lowe, A.~J., Smyth, A.~K., Atkins, K., Avery, R., Belbin, L., Brown, N.,
  Budden, A.~E., Gioia, P., Guru, S., Hardie, M., Hirsch, T., Hobern, D.,
  La~Salle, J., Loarie, S.~R., Miles, M., Milne, D., Nicholls, M., Rossetto,
  M., Smits, J., Sparrow, B., Terrill, G., Turner, D., and Wardle, G.~M.}
\newblock Publish openly but responsibly.
\newblock {\em Science 357}, 6347 (July 2017), 141--141.
\newblock Publisher: American Association for the Advancement of Science.

\bibitem{marzoev_towards_2019}
{\sc Marzoev, A., Araújo, L.~T., Schwarzkopf, M., Yagati, S., Kohler, E.,
  Morris, R., Kaashoek, M.~F., and Madden, S.}
\newblock Towards {Multiverse} {Databases}.
\newblock In {\em Proceedings of the {Workshop} on {Hot} {Topics} in
  {Operating} {Systems}\/} (New York, NY, USA, May 2019), {HotOS} '19,
  Association for Computing Machinery, pp.~88--95.

\bibitem{mbomaalternative}
{\sc Mboma, J. C.~A.}
\newblock Alternative livelihoods of forest edge communities around the gola
  rain forest national park in malema chiefdom, kailahun district.
\newblock {\em International Journal of Innovative Science, Engineering and
  Technology 8\/} (2021).

\bibitem{mcsherry2015scalability}
{\sc McSherry, F., Isard, M., and Murray, D.~G.}
\newblock Scalability! but at what $\{$COST$\}$?
\newblock In {\em 15th Workshop on Hot Topics in Operating Systems (HotOS
  XV)\/} (2015).

\bibitem{fairness-ml}
{\sc Mehrabi, N., Morstatter, F., Saxena, N., Lerman, K., and Galstyan, A.}
\newblock A survey on bias and fairness in machine learning.
\newblock {\em ACM Comput. Surv. 54}, 6 (jul 2021).

\bibitem{merali_computational_2010}
{\sc Merali, Z.}
\newblock Computational science: ...{Error}.
\newblock {\em Nature 467}, 7317 (Oct. 2010), 775--777.

\bibitem{ecocode}
{\sc Mislan, K., Heer, J.~M., and White, E.~P.}
\newblock Elevating the status of code in ecology.
\newblock {\em Trends in Ecology \& Evolution 31}, 1 (2016), 4--7.

\bibitem{moussa_scalable_2021}
{\sc Moussa, R.}
\newblock Scalable analytics of air quality batches with {Apache} {Spark} and
  {Apache} {Sedona}.
\newblock In {\em Proceedings of the 15th {ACM} {International} {Conference} on
  {Distributed} and {Event}-based {Systems}\/} (New York, NY, USA, June 2021),
  {DEBS} '21, Association for Computing Machinery, pp.~154--159.

\bibitem{murray_ciel_2011}
{\sc Murray, D.~G., Schwarzkopf, M., Smowton, C., Smith, S., Madhavapeddy, A.,
  and Hand, S.}
\newblock {CIEL}: a universal execution engine for distributed data-flow
  computing.
\newblock In {\em Proceedings of the 8th {USENIX} conference on {Networked}
  systems design and implementation\/} (USA, Mar. 2011), {NSDI}'11, USENIX
  Association, pp.~113--126.

\bibitem{mutanga2019google}
{\sc Mutanga, O., and Kumar, L.}
\newblock Google earth engine applications.
\newblock {\em Remote sensing 11}, 5 (2019), 591.

\bibitem{novella_container-based_2019}
{\sc Novella, J.~A., Emami~Khoonsari, P., Herman, S., Whitenack, D., Capuccini,
  M., Burman, J., Kultima, K., and Spjuth, O.}
\newblock Container-based bioinformatics with {Pachyderm}.
\newblock {\em Bioinformatics 35}, 5 (Mar. 2019), 839--846.

\bibitem{nugent2018inaturalist}
{\sc Nugent, J.}
\newblock inaturalist: citizen science for 21st-century naturalists.
\newblock {\em Science Scope 41}, 7 (2018), 12--15.

\bibitem{parsons_unlocking_2023}
{\sc Parsons, A., and Powell-Smith, A.}
\newblock Unlocking the value of fragmented public data, 2023.
\newblock
  \url{https://research.mysociety.org/publications/unlocking-fragmented-data}.

\bibitem{paszke_pytorch_2019}
{\sc Paszke, A., Gross, S., Massa, F., Lerer, A., Bradbury, J., Chanan, G.,
  Killeen, T., Lin, Z., Gimelshein, N., Antiga, L., Desmaison, A., Köpf, A.,
  Yang, E., DeVito, Z., Raison, M., Tejani, A., Chilamkurthy, S., Steiner, B.,
  Fang, L., Bai, J., and Chintala, S.}
\newblock {PyTorch}: an imperative style, high-performance deep learning
  library.
\newblock In {\em Proceedings of the 33rd {International} {Conference} on
  {Neural} {Information} {Processing} {Systems}}, no.~721. Curran Associates
  Inc., Red Hook, NY, USA, Dec. 2019, pp.~8026--8037.

\bibitem{peterson_design_2006}
{\sc Peterson, L., and Roscoe, T.}
\newblock The design principles of {PlanetLab}.
\newblock {\em ACM SIGOPS Operating Systems Review 40}, 1 (Jan. 2006), 11--16.

\bibitem{richards_are_2017}
{\sc Richards, P., Arima, E., VanWey, L., Cohn, A., and Bhattarai, N.}
\newblock Are {Brazil}'s {Deforesters} {Avoiding} {Detection}?
\newblock {\em Conservation Letters 10}, 4 (2017), 470--476.
\newblock \_eprint: https://onlinelibrary.wiley.com/doi/pdf/10.1111/conl.12310.

\bibitem{saeedan2022}
{\sc Saeedan, M., and Eldawy, A.}
\newblock Spatial parquet: A column file format for geospatial data lakes.
\newblock In {\em Proceedings of the 30th International Conference on Advances
  in Geographic Information Systems\/} (New York, NY, USA, 2022), SIGSPATIAL
  '22, Association for Computing Machinery.

\bibitem{sarkar2022like}
{\sc Sarkar, A., Gordon, A.~D., Negreanu, C., Poelitz, C., Ragavan, S.~S., and
  Zorn, B.}
\newblock What is it like to program with artificial intelligence?
\newblock {\em arXiv preprint arXiv:2208.06213\/} (2022).

\bibitem{doi:10.1177/19400829211014740}
{\sc Sarkar, D., and Chapman, C.~A.}
\newblock The smart forest conundrum: Contextualizing pitfalls of sensors and
  ai in conservation science for tropical forests.
\newblock {\em Tropical Conservation Science 14\/} (2021), 19400829211014740.

\bibitem{seth2022technology}
{\sc Seth, A.}
\newblock {\em Technology and (dis) empowerment: A Call to Technologists}.
\newblock Emerald Publishing Limited, 2022.

\bibitem{simmhan_survey_2005}
{\sc Simmhan, Y.~L., Plale, B., and Gannon, D.}
\newblock A survey of data provenance in e-science.
\newblock {\em ACM SIGMOD Record 34}, 3 (Sept. 2005), 31--36.

\bibitem{slough_satellite-based_2021}
{\sc Slough, T., Kopas, J., and Urpelainen, J.}
\newblock Satellite-based deforestation alerts with training and incentives for
  patrolling facilitate community monitoring in the {Peruvian} {Amazon}.
\newblock {\em Proceedings of the National Academy of Sciences 118}, 29 (July
  2021), e2015171118.

\bibitem{rspet}
{\sc Society, T.~R.}
\newblock From privacy to partnership.
\newblock Available at \url{royalsociety.org/privacy-enhancing-technologies}.

\bibitem{tigerpoach}
{\sc Storm, D.}
\newblock Cyber-poaching: Hacking gps collar data to track and kill endangered
  tigers.
\newblock Tech. rep., October 2013.
\newblock
  https://www.computerworld.com/article/2475200/cyber-poaching--hacking-gps-collar-data-to-track-and-kill-endangered-tigers.html.

\bibitem{cowg}
{\sc Swinfield, T., and Balmford, A.}
\newblock Cambridge carbon impact: Evaluating carbon credit claims and
  co-benefits.
\newblock Tech. rep., March 2023.

\bibitem{swinfield2023realising}
{\sc Swinfield, T., Balmford, A., Coomes, D., Madhavapeddy, A., and Keshav, S.}
\newblock Realising the social value of impermanent carbon credits.
\newblock {\em Nature Climate Change 13\/} (2023), 1172--1178.

\bibitem{tan2020scaling}
{\sc Tan, X., Zhou, M., and Fitzgerald, B.}
\newblock Scaling open source communities: An empirical study of the linux
  kernel.
\newblock In {\em Proceedings of the ACM/IEEE 42nd International Conference on
  Software Engineering\/} (2020), pp.~1222--1234.

\bibitem{tang_drone_2015}
{\sc Tang, L., and Shao, G.}
\newblock Drone remote sensing for forestry research and practices.
\newblock {\em Journal of Forestry Research 26}, 4 (Dec. 2015), 791--797.

\bibitem{tiancusz2020}
{\sc Tian, J., Di, S., Zhao, K., Rivera, C., Fulp, M.~H., Underwood, R., Jin,
  S., Liang, X., Calhoun, J., Tao, D., and Cappello, F.}
\newblock Cusz: An efficient gpu-based error-bounded lossy compression
  framework for scientific data.
\newblock In {\em Proceedings of the ACM International Conference on Parallel
  Architectures and Compilation Techniques\/} (2020), PACT '20, Association for
  Computing Machinery, p.~3–15.

\bibitem{tollefson_humans_2019}
{\sc Tollefson, J.}
\newblock Humans are driving one million species to extinction.
\newblock {\em Nature 569}, 7755 (May 2019), 171--171.

\bibitem{tollefson_carbon_2022}
{\sc Tollefson, J.}
\newblock Carbon emissions hit new high: warning from {COP27}.
\newblock {\em Nature\/} (Nov. 2022).
\newblock Bandiera\_abtest: a Cg\_type: News Publisher: Nature Publishing Group
  Subject\_term: Climate change, Policy, Climate sciences, Databases.

\bibitem{trolliet2014use}
{\sc Trolliet, F., Vermeulen, C., Huynen, M.-C., and Hambuckers, A.}
\newblock Use of camera traps for wildlife studies: a review.
\newblock {\em Biotechnologie, Agronomie, Soci{\'e}t{\'e} et Environnement 18},
  3 (2014).

\bibitem{tulloch2018decision}
{\sc Tulloch, A.~I., Auerbach, N., Avery-Gomm, S., Bayraktarov, E., Butt, N.,
  Dickman, C.~R., Ehmke, G., Fisher, D.~O., Grantham, H., Holden, M.~H.,
  et~al.}
\newblock A decision tree for assessing the risks and benefits of publishing
  biodiversity data.
\newblock {\em Nature ecology \& evolution 2}, 8 (2018), 1209--1217.

\bibitem{tmf}
{\sc Vancutsem, C., Achard, F., Pekel, J.-F., Vieilledent, G., Carboni, S.,
  Simonetti, D., Gallego, J., Aragão, L. E. O.~C., and Nasi, R.}
\newblock Long-term (1990–2019) monitoring of forest cover changes in the
  humid tropics.
\newblock {\em Science Advances 7}, 10 (2021), eabe1603.

\bibitem{automated-decision-making}
{\sc Waldman, A.~E.}
\newblock Power, process, and automated decision-making.
\newblock {\em Fordham L. Rev. 88\/} (2019), 613.

\bibitem{weiss_global_2020}
{\sc Weiss, D.~J., Nelson, A., Vargas-Ruiz, C.~A., Gligorić, K., Bavadekar,
  S., Gabrilovich, E., Bertozzi-Villa, A., Rozier, J., Gibson, H.~S., Shekel,
  T., Kamath, C., Lieber, A., Schulman, K., Shao, Y., Qarkaxhija, V., Nandi,
  A.~K., Keddie, S.~H., Rumisha, S., Amratia, P., Arambepola, R., Chestnutt,
  E.~G., Millar, J.~J., Symons, T.~L., Cameron, E., Battle, K.~E., Bhatt, S.,
  and Gething, P.~W.}
\newblock Global maps of travel time to healthcare facilities.
\newblock {\em Nature Medicine 26}, 12 (Dec. 2020), 1835--1838.

\bibitem{whitehead_precision_2011}
{\sc Whitehead, N., and Fit-Florea, A.}
\newblock Precision and {Performance}: {Floating} {Point} and {IEEE} 754
  {Compliance} for {NVIDIA} {GPUs}.

\bibitem{wulder2022fifty}
{\sc Wulder, M.~A., Roy, D.~P., Radeloff, V.~C., Loveland, T.~R., Anderson,
  M.~C., Johnson, D.~M., Healey, S., Zhu, Z., Scambos, T.~A., Pahlevan, N.,
  et~al.}
\newblock Fifty years of landsat science and impacts.
\newblock {\em Remote Sensing of Environment 280\/} (2022), 113195.

\bibitem{yang_two_2015}
{\sc Yang, J.-H., and Chan, B. P.-L.}
\newblock Two new species of the genus {Goniurosaurus} ({Squamata}: {Sauria}:
  {Eublepharidae}) from southern {China}.
\newblock {\em Zootaxa 3980}, 1 (June 2015), 67--80.
\newblock Number: 1.

\bibitem{ethics-remote-sensing}
{\sc York, N.~D., Pritchard, R., Sauls, L.~A., Enns, C., and Foster, T.}
\newblock Justice and ethics in conservation remote sensing: Current discourses
  and research needs.
\newblock {\em Biological Conservation 287\/} (2023), 110319.

\bibitem{zeldovich2008securing}
{\sc Zeldovich, N., Boyd-Wickizer, S., and Mazieres, D.}
\newblock Securing distributed systems with information flow control.
\newblock In {\em NSDI\/} (2008), vol.~8, pp.~293--308.

\bibitem{zeldovich_securing_2008}
{\sc Zeldovich, N., Boyd-Wickizer, S., and Mazières, D.}
\newblock Securing distributed systems with information flow control.
\newblock In {\em Proceedings of the 5th {USENIX} {Symposium} on {Networked}
  {Systems} {Design} and {Implementation}\/} (USA, Apr. 2008), {NSDI}'08,
  USENIX Association, pp.~293--308.

\end{thebibliography}
}

\end{document}